\documentclass[cits]{PoS}
\usepackage{epsfig}

\title{The chiral condensate on 2--flavor staggered configurations from an overlap operator}

\ShortTitle{The chiral condensate on 2--flavor staggered configurations from an overlap operator}

\author{\speaker{Anna Hasenfratz} and Roland Hoffmann\\
        Department of Physics, University of Colorado, Boulder, CO-80309-390\\
        E-mail: \email{anna@eotvos.colorado.edu}}

\abstract{We measure the low lying eigenmodes of an overlap Dirac operator on 2--flavor
staggered configurations. By comparing the eigenmode distribution to the predictions of Random
Matrix Theory we test to what accuracy staggered configurations
 describe  continuum QCD.  The
agreement between the numerical data and RMT implies that at our
quark mass values the lattice artifacts of the staggered configurations 
are comparable to overlap configurations.
We identify the overlap valence mass that best matches the staggered sea quarks and  predict the value of the infinite volume 2--flavor
chiral condensate.}

\FullConference{XXIVth International Symposium on Lattice Field Theory\\
                July 23-28, 2006\\
                Tucson, Arizona, USA}

\begin{document}

\section{Introduction}

Mixed action simulations became popular in recent years as they combine
the simulation advantages of a simple sea quark action with the exact
or near exact chiral symmetry of overlap or domain wall valence quarks.
The price to pay, in addition to an internal inconsistency (unitarity
violation), is the complication in the analysis. One option is to
derive and use partially quenched mixed action chiral perturbative
formulae. Alternatively, one can match the parameters
of the valence and sea quark actions as well as possible and deal
with any remaining difference as part of the lattice artifacts. This
approach is useful if the chiral perturbative formulae do not exist
or the numerical data does not allow the fitting of all the parameters,
or if one desires more insight into the physics contained in a particular
set of gauge configurations.

The effectiveness of the latter approach was illustrated in Refs.
\cite{Hasenfratz:2006nw,Hoffmann:2006XX},
where we showed that,
at least within the 2--dimensional Schwinger model, mixed action simulations
with overlap valence quarks on rooted or unrooted staggered sea quark configurations
reproduce the full dynamical overlap results if the overlap mass is tuned appropriately.
In this work we report our first results along the same lines in 4--dimensional
2--flavor QCD. We show that configurations generated with 2--flavor
staggered quarks at a single lattice spacing but at four different
quark masses are consistent with 2--flavor QCD configurations, at
least for the three heavier masses. We also show how the topological
charge distributions can be used to determine the best overlap valence
matching masses and that with these mass values the data sets predict
a consistent value for the chiral condensate of 2--flavor QCD. All details can
be found in Ref. \cite{THEPAPER}.

\section{Strategy and Simulation setup\label{sec:Strategy-and-Simulation}}

Our sea quark action is the 2--flavor Asqtad staggered action \cite{Orginos:1999cr,Bernard:2001av,Aubin:2004wf}.
We have generated four configurations sets, each consisting of 400-500
$12^{4}$ lattices at a lattice spacing of about $a=0.13\,$fm. 
The details of the sets are summarized in Table \ref{cap:Parameters-of-the}.
The level of
taste breaking, the ratios of the heaviest and lightest pion masses,
is approximated from corresponding 2+1 flavor results \cite{Bernard:2001av}.
The last column lists the separation of the configurations in terms
of unit length molecular dynamics trajectories. 	 

Our valence action is an overlap action based on an improved Wilson kernel 
on HYP smeared links. This action was used in recent overlap 
simulations  \cite{DeGrand:2000tf,DeGrand:2003in}.

To investigate if the rooted staggered gauge ensembles are 
consistent with
two--flavor continuum QCD one has to consider observables
that are sensitive to the vacuum and do not depend strongly on the
valence quark mass. Spectral quantities are not appropriate, but the
low lying infrared eigenmodes of the massless valence Dirac operator
offer a good choice. Another quantity which we will consider is the topological charge of the configurations. 
On each configurations we ask what valence quark mass matches the staggered configurations the best, i.e. what  valence mass minimizes the lattice artifacts, the difference between lattice data and continuum QCD. 

{\small
\TABULAR[t]{|c|c|c|c|c|c|}{
\hline 
$\ $Set$\ $&
$\quad\beta\quad$&
$\ am_{st}\ $&
$r_{0}/a$ \cite{Sommer:1993ce,Hasenfratz:2001tw} &
$\ $Taste breaking$\ $&
$\ $Time separation$\ $\tabularnewline
\hline
\hline 
\textbf{L}&7.18&0.01&3.84(6)&60\%&5\tabularnewline
\hline
\textbf{M}&7.20&0.02&3.82(3)&34\%&5\tabularnewline
\hline 
\textbf{H}&7.22&0.03&3.60(4)&24\%&10\tabularnewline
\hline
\textbf{E}&7.24&0.04&3.64(3)&18\%&15\tabularnewline
\hline}
{Parameters of the $n_{f}\!=\!2$ staggered background configurations. The molecular dynamics time separations
between the configurations
reflect the autocorrelation of the topological charge. \label{cap:Parameters-of-the}}}

\section{Eigenvalues of the Dirac Operator and Random Matrix Theory\label{sec:Eigenvalues-of-the} }

Random matrix theory (RMT) captures the universal
chiral properties of QCD and predicts the distribution of the physical
(infrared) eigenvalues of the massless Dirac operator in the $\epsilon$--regime.
The predictions are given in fixed topological sector $\nu$ and depend
on the low energy constant $\Sigma$, the infinite volume chiral condensate.
The distribution of the (microscopically
rescaled) $n^{\mathrm{th}}$ eigenmode $\lambda\Sigma V$ is given
as \begin{equation}
P_{\nu,n}(\lambda\Sigma V)=\Lambda_{\nu,n}(m\Sigma V;n_{f})\;,\label{RMT-pred}\end{equation}
where $m$ is the quark mass of the configurations (sea quark mass), which
in our case is an overlap quark mass that corresponds to the background
configurations that were generated by staggered quarks, i.e. $m$
is the matching quark mass as described in Sect. \ref{sec:Strategy-and-Simulation}.
The value of $m$ is not known a priori and therefore
$\Lambda_{\nu,n}$ depends on two variables, $M=m\Sigma V$ and $\Sigma.$ We
fit the measured eigenvalue distribution to random matrix theory at
fixed $M$ and predict the chiral condensate $\Sigma$. The systematic
deviation of the data from the RMT prediction of Eq.(\ref{RMT-pred})
characterizes the lattice artifacts, both from discretization errors
and from the non-locality of the action. This deviation is the measure
of consistency between the lattice action and continuum QCD and replaces
the residue used in Ref.\cite{Hasenfratz:2006nw} for the same purpose.
If the rooting procedure is correct, it should scale to zero as the
continuum limit is approached at fixed physical (matching) quark mass,
assuming the simulations are done in the region where the RMT predictions
are valid.

To fit the cumulative distributions, we use the  Kolmogorov-Smirnov (KS) test
that minimizes $D_{\mathrm{max}}^{2}$, the maximal deviation between the
measured and the predicted cumulative
distributions \cite{Bietenholz:2003mi,DeGrand:2006uy,DeGrand:2006nv}.
An advantage of the KS test is that there is an explicit and simple
form for the confidence level of the fit. For a given sample length
this quality factor $Q_{\mathrm{KS}}$
is a monotonically decreasing function of $D_{\mathrm{max}}$ that gives the probability
that the measured distribution is consistent with the analytical one.
The KS fit maximizes the quality factor $Q_{\mathrm{KS}}$ or the
product of quality factors if more than one distribution is used.

However, $Q_{\mathrm{KS}}$ will go to zero exponentially with increasing
statistics if the measured distribution is not \emph{exactly} described
by the analytic form. In any lattice calculations there are lattice
artifacts and finite volume effects, so the analytic form is never
exactly reproduced, the quality factor vanishes as the numerical statistics
increases. In the following we fit our data by maximizing
the quality factor (or products of quality factors) according to
the KS test, but describe the goodness of the fit by the value $D_{\mathrm{max}}$
itself to enable a comparison with other results.

\EPSFIGURE[t]{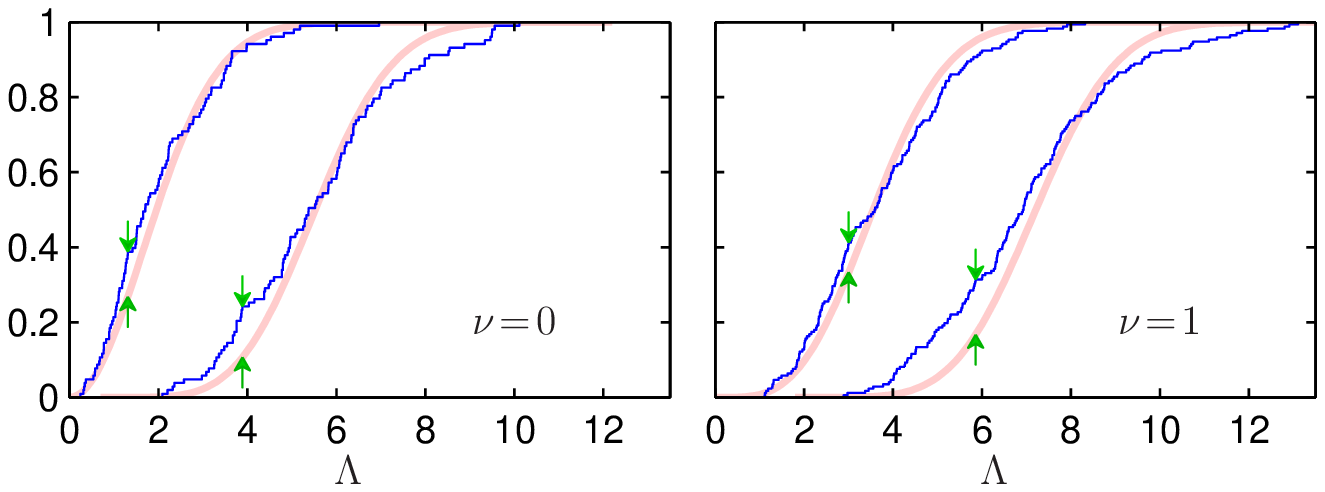,width=115mm}{
RMT predictions of the cumulative distribution of the two lowest
eigenmodes in the $\nu=0$ and $1$ sectors of the \textbf{M} set
at $M\!=\!13.5$ (see below). The fit uses
only the first mode in each topological sector. Arrows
indicate the maximal deviation between the data set and the analytical
predictions.\label{cap:RMT-fit}}
Fig. \ref{cap:RMT-fit} shows a typical 
fit of the cumulative distribution for the \textbf{M} ($am_{st}=0.02)$
data set at $M\!=\!13.5$. The left panel corresponds to the $\nu=0$, 
the right panel
to the $\nu=1$ sector. Only the first modes of the $\nu=0$ and 1 topological
sectors are included in the fit. In addition to the two fitted modes
we also show the non-fitted second modes in the same topological sectors.
$D_{\mathrm{max}}$ is almost a factor of two smaller for the $n=1$,
$\nu=1$ mode, but not significantly worse for the non-fitted modes
than for the fitted $n=1$, $\nu=0$ mode.

In Fig.\ref{DM} we plot the maximal deviations $D_{\mathrm{max}}$
as a function of the RMT parameter $M$. Evidently the quality of
the fit is not very sensitive to the parameter $M$. While small values
are disfavored, larger values are almost equally probable. Contrary
to our original hopes the eigenmode distributions cannot be used to
define a matching mass, it defines only a range of acceptable values.

\EPSFIGURE[!h]{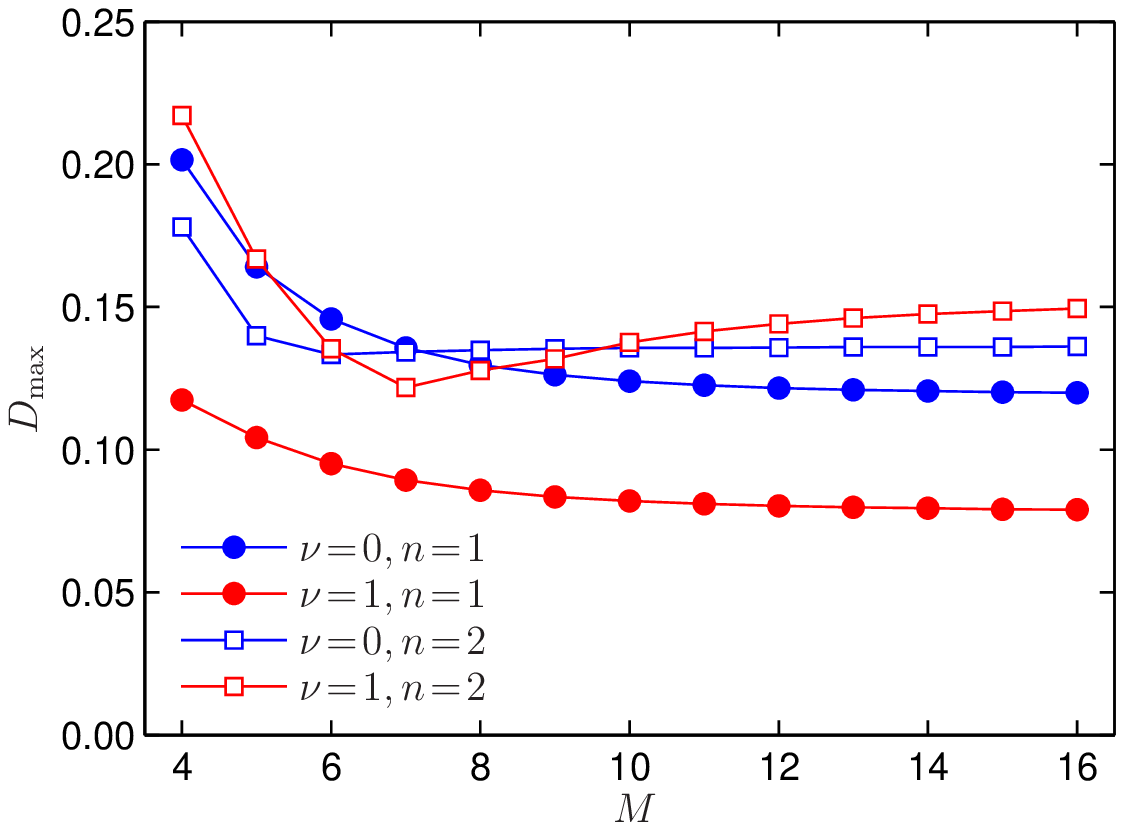,width=75mm}{
$D_{\mathrm{max}}$ as the function of $M$
for the \textbf{M} data set. The fit uses the first modes of the
$\nu=0$ and 1 sectors (filled points).\label{DM}}

Result of the fit are similar for the other three data sets. The
upper panels of Figure \ref{cap:Sigma for all} show $D_{\mathrm{max}}$
for the fitted modes, and the dependence on the staggered mass is
obvious. $D_{\mathrm{max}}$ is significantly lower at the heaviest
\textbf{E} data set than for the lightest \textbf{L} one, with the
intermediate mass sets lying in between. This behavior is expected
since at finite lattice spacing a smaller staggered mass leads to
increased taste symmetry breaking (Table \ref{cap:Parameters-of-the}),
it differs more from the flavor symmetric valence quark sector. With
decreasing lattice spacing at fixed physical quark mass this deviation
should decrease and eventually vanish in the continuum limit.

At each $M$ value the fit predicts $\Sigma V/a$ and using
$r_{0}/a$ from Table \ref{cap:Parameters-of-the}
this can be converted to physical units as shown on the lower panels
of Fig.\ref{cap:Sigma for all}. The corresponding  overlap
mass values $m$ are shown along the upper border of the figure.
In order to predict the chiral condensate we have to find an
independent quantity that predicts the matching valence quark mass.
The topological charge distribution is a possible choice as we will
discuss in Sect. \ref{sec:Topology}.

\EPSFIGURE[t]{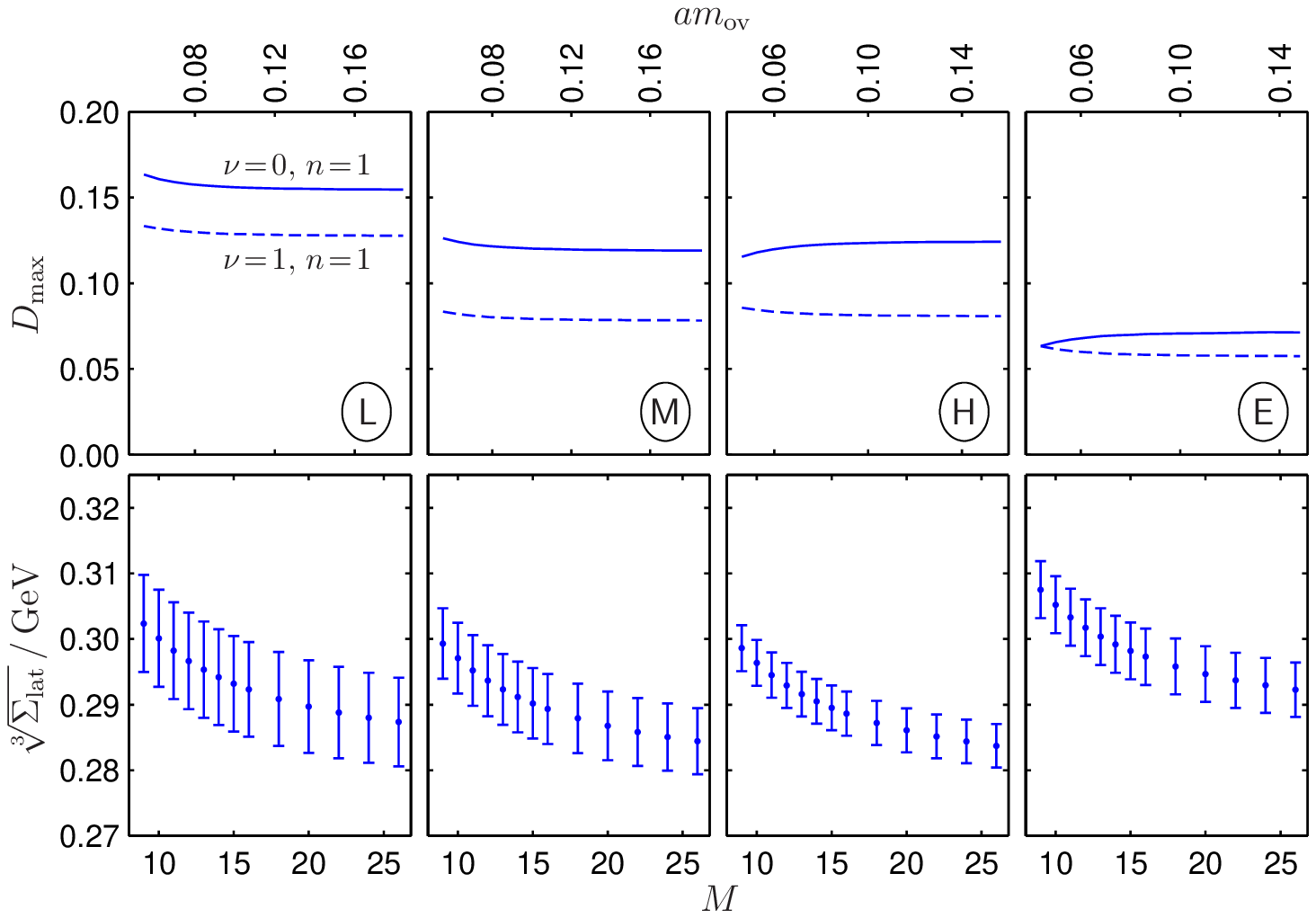,width=115mm}{
$D_{\mathrm{max}}$ and $\Sigma^{1/3}$ in GeV as the function of
$M=m\Sigma V$ for all four data sets. \label{cap:Sigma for all}}
{\small
\TABULAR[b]{|c|c|c|c|c|r|c|c|}{
\hline 
$\ $Set$\ $&
$\ M=m\Sigma V\ $&
$\ \Sigma^{1/3}$ / MeV$\ $&
$am$&
$\ \nu\ $&
$\ N\ $&
$\ D_{{\rm max}}\ $&
$\ Q_{{\rm KS}}\ $\tabularnewline
\hline
\hline\textbf{L}&12.7(2.0)&295.7(7.0)&0.083(4)(14)&0&89&0.157&0.022\tabularnewline
\hline\multicolumn{4}{|c|}{}&1&144&0.130&0.014\tabularnewline
\hline \textbf{M}&13.5(2.8)&291.7(4.1)&0.090(3)(19)&0&103&0.121&0.092\tabularnewline
\hline\multicolumn{4}{|c|}{}&1&172&0.080&0.217\tabularnewline
\hline \textbf{H}&16.9(1.9)&288.0(5.4)&0.098(4)(12)&0&87&0.123&0.132\tabularnewline
\hline \multicolumn{4}{|c|}{}&1&178&0.081&0.181\tabularnewline
\hline \textbf{E}&22.6(4.3)&293.5(4.1)&0.127(4)(24)&0&85&0.071&0.768\tabularnewline
\hline \multicolumn{4}{|c|}{}&1&193&0.058&
0.022\tabularnewline
\hline}
{Results of the RMT fit to the lowest eigenmodes in the $\nu=0,1$
sectors. For the determination of $M$
see Sect. \ref{sec:Topology}. The first error on the matching overlap
mass $am$ is due to the uncertainty of $\Sigma$ only, while the
second one takes into account both the errors of $\Sigma$ and $M$.
\label{cap:Results}}}
In Table \ref{cap:Results} we list the number
of configurations,
the $D_{\mathrm{max}}$ values of the RMT fit at specific $M=m\Sigma V$
values and the corresponding quality factors. The $D_{\mathrm{max}}$ values 
can be compared to those from Ref.  \cite{DeGrand:2006nv}. 
That work uses dynamical overlap configurations
at similar physical volumes at slightly coarser lattice spacing.
Using the same fitting strategy as ours  they find 
 $D_{max} = 0.11\!\sim\!0.20$. In view
of these numbers we can conclude that, as far as the Dirac operator
eigenmode distribution is concerned, the rooted staggered action configurations
do not show larger lattice artifacts than the overlap ones.

\section{Topology\label{sec:Topology}}


Since we have sufficient statistics, over 400 approximately independent
configurations at each coupling value on not too large volumes ($12^{4}$
or about 6 fm$^{4}$), we can study the topological charge distribution.
Following the discussion of Refs. \cite{Leutwyler:1992yt,Verbaarschot:2000dy,Durr:2001ty},
we write the probability of encountering a charge $\pm\nu$ configuration
in the dynamical ensemble as $
P_{\nu}=Z_{\nu}(m\Sigma V)Q_{\nu}(\sigma)$.
Here $Q_{\nu}$ is the  \emph{quenched} probability of a charge $\pm\nu$
configuration, expected to be Gaussian up to  $1/V$  corrections,   while $Z_{\nu}$ describes the suppression due to the
fermionic determinant. The fermionic
suppression factor has been calculated both within chiral perturbation
theory and the random matrix model \cite{Leutwyler:1992yt,Verbaarschot:2000dy}.
Thus the charge probability
distribution $P_{\nu}$ depends on two variables, $M=m\Sigma V$ and
$\sigma$. The latter can be determined from the quenched topological
susceptibility, so a one parameter fit to the topological charge distribution
data predicts $M$, which we list in Table
\ref{cap:Results}.

\section{The chiral condensate and matching masses\label{sec:The-chiral-condensate}}

With the $M=m\Sigma V$ values predicted from the topological charge
distribution we are now able to extract the physical value of the
chiral condensate. Combining the $M$ values with $r_{0}/a$ from
Table \ref{cap:Parameters-of-the} we find that all four configuration
sets predict a consistent value for the $\Sigma$ condensate, as listed
in Table \ref{cap:Results}. The only sign that the light \textbf{L}
set differs from the RMT prediction more than the other mass values
is the larger error of the predicted condensate. The value we obtain,
\begin{equation}
\Sigma_{\mathrm{lat}}^{1/3}=291(5)\;\mathrm{MeV}\;,\label{eq:Sigma}\end{equation}
is the lattice condensate. It is consistent with predictions obtained on overlap dynamical
configurations \cite{DeGrand:2006nv},
further supporting our observation that the rooted
staggered configurations are QCD like, the non-local terms of the
action can be simply taken into account as lattice artifacts.

To connect the value of the condensate to a more conventional scheme, like $\overline{\mathrm{MS}}$
at 2$\,$GeV, one needs the corresponding renormalization factor $Z_{s}$.
Such a factor should be calculated non-perturbatively on the staggered
configurations with our specific valence Dirac operator. We have not
done this calculation yet but similar ones exist
\cite{DeGrand:2006uy,DeGrand:2006nv,DeGrand:2005af}.
$Z_{s}$ seems to be largely independent of the detailed properties
of the background configurations and we estimate its value to be
$Z_{s}\geq0.9$,
which will lower $\Sigma_{\mathrm{lat}}^{1/3}$ by 3\%. In
addition, there is a finite volume correction to the condensate
that could lower its value further \cite{DeGrand:2006nv}.
These effects will have to be investigated but they are beyond the scope of
the present work. 

Combining $M$ and $\Sigma$ we get the values listed in the ''$am$''
column of Table \ref{cap:Results}. These matching masses are not
only surprisingly large but they do not depend linearly on the staggered
masses. While the staggered quark mass changes a factor of four between
the lightest and heaviest data sets, the matching overlap masses change
only 50\% . This is similar to what we observed in the Schwinger model
\cite{Hasenfratz:2006nw}. The matching valence masses show an overall
shift compared to the staggered sea mass values. In addition at very
small sea quark masses, where the matching breaks down, the valence
quark masses are largely independent of the sea quark mass values.
This is illustrated in Fig. 3 of Ref. \cite{Hasenfratz:2006nw}. Such
behavior implies that staggered configurations at small quark masses
are not necessarily closer to chiral continuum QCD than the heavier
mass configurations. All the computational efforts creating light
configurations might be in vain, creating only configurations with
larger lattice artifacts. This might not be a problem when the data
is analyzed with the whole machinery of staggered partially quenched
chiral perturbation theory but should be considered when individual
configuration sets are analyzed in mixed action simulations. Of course
this is only a lattice artifact and any such effect will disappear
as the continuum limit is approached. 

\section{Conclusion}

We have studied the properties of the rooted staggered action in a
mixed action simulation using overlap valence quarks. By comparing
physical quantities that are independent of the valence quark mass
to continuum QCD predictions we identified lattice artifacts and
studied their dependence on the sea quark masses.
In this work we considered the eigenvalue distribution of the massless
Dirac operator and the distribution of the topological charge. We
compared the former to the universal predictions of random matrix
theory and found that the systematic deviation of the data from the
predictions were comparable to dynamical overlap simulations.
Using the topological charge
distribution we could identify the matching overlap valence quark
mass value which best describes the staggered configurations. We found
these matching values to be fairly large and their dependence on the
staggered sea mass values indicate a finite offset, in addition to
a linear mass renormalization factor between the valence and sea mass
values. With the use of this matching mass we extracted the value
of the chiral scalar condensate. We found that the predictions from
all of our staggered configuration sets were consistent. These findings
indicate that at our lattice spacing, $a\!\approx\!0.13\,$fm, and
with not very light sea quarks the rooted staggered lattice configurations
have lattice artifacts similar to other lattice actions, the non-local
terms arising from the rooting procedure can be simply considered
as part of the cutoff effects. In order to show that these non-local
terms indeed become irrelevant in the continuum limit the calculation
have to be repeated at different lattice spacings and the scaling
of the lattice artifacts should be investigated. It would also be important
to study in a similar manner the lattice artifacts of other observables.

{\renewcommand{\baselinestretch}{0.87}
\bibliography{lattice}

\providecommand{\href}[2]{#2}\begingroup\raggedright\begin{thebibliography}{10}

\bibitem{Hasenfratz:2006nw}
A.~Hasenfratz and R.~Hoffmann {\em Phys. Rev.} {\bf D74} (2006) 014511,
  [\href{http://xxx.lanl.gov/abs/hep-lat/0604010}{{\tt hep-lat/0604010}}].

\bibitem{Hoffmann:2006XX}
A.~Hasenfratz and R.~Hoffmann {\em PoS} {\bf LAT2006} (2006) 212,
  [\href{http://xxx.lanl.gov/abs/hep-lat/0609030}{{\tt hep-lat/0609030}}].

\bibitem{THEPAPER}
A.~Hasenfratz and R.~Hoffmann, {\it Mixed action simulations on staggered
  background; interpretation and result for the {2-flavor} {QCD} chiral
  condensate},  \href{http://xxx.lanl.gov/abs/hep-lat/0609067}{{\tt
  hep-lat/0609067}}.

\bibitem{Orginos:1999cr}
{\bf MILC} Collaboration, K.~Orginos, D.~Toussaint, and R.~L. Sugar {\em Phys.
  Rev.} {\bf D60} (1999) 054503,
  [\href{http://xxx.lanl.gov/abs/hep-lat/9903032}{{\tt hep-lat/9903032}}].

\bibitem{Bernard:2001av}
C.~W. Bernard {\em et~al.} {\em Phys. Rev.} {\bf D64} (2001) 054506,
  [\href{http://xxx.lanl.gov/abs/hep-lat/0104002}{{\tt hep-lat/0104002}}].

\bibitem{Aubin:2004wf}
C.~Aubin {\em et~al.} {\em Phys. Rev.} {\bf D70} (2004) 094505,
  [\href{http://xxx.lanl.gov/abs/hep-lat/0402030}{{\tt hep-lat/0402030}}].

\bibitem{DeGrand:2000tf}
{\bf MILC} Collaboration, T.~A. DeGrand {\em Phys. Rev.} {\bf D63} (2001)
  034503, [\href{http://xxx.lanl.gov/abs/hep-lat/0007046}{{\tt
  hep-lat/0007046}}].

\bibitem{DeGrand:2003in}
{\bf MILC} Collaboration, T.~A. DeGrand {\em Phys. Rev.} {\bf D69} (2004)
  014504, [\href{http://xxx.lanl.gov/abs/hep-lat/0309026}{{\tt
  hep-lat/0309026}}].

\bibitem{Sommer:1993ce}
R.~Sommer {\em Nucl. Phys.} {\bf B411} (1994) 839--854,
  [\href{http://xxx.lanl.gov/abs/hep-lat/9310022}{{\tt hep-lat/9310022}}].

\bibitem{Hasenfratz:2001tw}
A.~Hasenfratz, R.~Hoffmann, and F.~Knechtli {\em Nucl. Phys. Proc. Suppl.} {\bf
  106} (2002) 418--420, [\href{http://xxx.lanl.gov/abs/hep-lat/0110168}{{\tt
  hep-lat/0110168}}].

\bibitem{Bietenholz:2003mi}
W.~Bietenholz, K.~Jansen, and S.~Shcheredin {\em JHEP} {\bf 07} (2003) 033,
  [\href{http://xxx.lanl.gov/abs/hep-lat/0306022}{{\tt hep-lat/0306022}}].

\bibitem{DeGrand:2006uy}
T.~DeGrand, R.~Hoffmann, S.~Schaefer, and Z.~Liu {\em Phys. Rev.} {\bf D74}
  (2006) 054501, [\href{http://xxx.lanl.gov/abs/hep-th/0605147}{{\tt
  hep-th/0605147}}].

\bibitem{DeGrand:2006nv}
T.~DeGrand, Z.~Liu, and S.~Schaefer, {\it Quark condensate in two-flavor
  {QCD}},  \href{http://xxx.lanl.gov/abs/hep-lat/0608019}{{\tt
  hep-lat/0608019}}.

\bibitem{Leutwyler:1992yt}
H.~Leutwyler and A.~Smilga {\em Phys. Rev.} {\bf D46} (1992) 5607--5632.

\bibitem{Verbaarschot:2000dy}
J.~J.~M. Verbaarschot and T.~Wettig {\em Ann. Rev. Nucl. Part. Sci.} {\bf 50}
  (2000) 343--410, [\href{http://xxx.lanl.gov/abs/hep-ph/0003017}{{\tt
  hep-ph/0003017}}].

\bibitem{Durr:2001ty}
S.~D{\"u}rr {\em Nucl. Phys.} {\bf B611} (2001) 281--310,
  [\href{http://xxx.lanl.gov/abs/hep-lat/0103011}{{\tt hep-lat/0103011}}].

\bibitem{DeGrand:2005af}
T.~A. DeGrand and Z.~Liu {\em Phys. Rev.} {\bf D72} (2005) 054508,
  [\href{http://xxx.lanl.gov/abs/hep-lat/0507017}{{\tt hep-lat/0507017}}].

\end{thebibliography}\endgroup
\bibliographystyle{JHEP}}

\end{document}